\documentclass[aps,showpacs,pre,onecolumn]{revtex4-1}

\usepackage[T1]{fontenc}
\usepackage[latin1]{inputenc}
\usepackage{amsmath}
\usepackage{amsthm}
\usepackage{amssymb}
\usepackage{subfigure}	
\usepackage{graphicx}
\usepackage{epstopdf}
\usepackage{color}
\usepackage[colorlinks=true,linkcolor=blue,citecolor=blue]{hyperref}
\usepackage{extramarks}
\usepackage{ae}
\usepackage{color}

\begin{document}
\date{}
\title{Rogue waves of Ultra-High Peak Amplitude:\\
A Mechanism for Reaching up to Thousand Times the Background Level}
\author{ Wen-Rong Sun}
\affiliation{School of Mathematics and Physics, University of Science\\
  and Technology Beijing, Beijing 100083, China}
\author{Lei Liu$^*$}
\affiliation{Beijing Computational Science Research Center, Beijing 100193, China}
\author{P.G. Kevrekidis}
\affiliation{Department of Mathematics and Statistics, University of
  Massachusetts, Amherst, Massachusetts 01003-4515, USA}
\affiliation{Mathematical Institute, University of Oxford, OX26GG, UK}


\begin{abstract}
We unveil a  mechanism enabling a fundamental rogue
wave, expressed by a rational  function of  fourth degree, to reach a
peak amplitude as high as a thousand times the background level in a
system of coupled nonlinear Schr\"{o}dinger equations involving both
incoherent
and coherent coupling terms with suitable coefficients. We obtain the exact explicit vector rational
solutions
using a Darboux-dressing transformation.  We show that {\it both} components
of such coupled equations can reach extremely
high amplitudes. The mechanism is confirmed in direct
numerical
simulations and its robustness confirmed upon noisy perturbations.
Additionally, we showcase the fact that extremely high peak-amplitude
vector fundamental rogue waves (of about 80 times the background
level) can
be excited even within a {\it chaotic background field}.
\end{abstract}

\maketitle

\newpage

\noindent\textbf{\large 1. Introduction}\\\hspace*{\parindent}

Rogue waves or waves of extreme amplitude
have been the subject of intense research activity
recently~\cite{1,2,3}. Within the
broad framework of the focusing
nonlinear Schr\"{o}dinger (NLS) equation, the analogous
physics of
light propagation
and hydrodynamic waves has led to a significant volume of corresponding
considerations
in nonlinear optics~\cite{xj1,xj2,xj21}. In $2007$, the concept of rogue waves in optics was first used to describe the rare, extreme fluctuations in the value of an optical field~\cite{4}. Since then the optical rogue waves have been generalized to describe many other processes in this area~\cite{41,pr1,pr2,pr3,pr4,pr5,pr6}.
The most common mathematical description of a rogue wave is the
Peregrine soliton, which can be expressed by a rational function of
second degree. This waveform presents a double spatio-temporal
algebraic localization on a finite continuous
background~\cite{5,xj21}. The Peregrine soliton is the simplest
prototype of
a fundamental rogue wave~\cite{xj1,xj2}. The dynamics of a fundamental rogue wave was observed
in numerous physical settings, including nonlinear fibers~\cite{xj2},
water wave tanks~\cite{nf2} and plasmas~\cite{nf3}.

Recent studies reveal that
in the presence of higher order effects and coupling between
components (in multi-component systems), more complex waveforms than the Peregrine
soliton can also arise~\cite{6,7,8,tt1,tt2}. For example, the
mathematical descriptions of fundamental rogue waves governed by the
Sasa-Satsuma (SS) and coupled SS models involve fourth-order
polynomial waveforms~\cite{6,tt1}. The orthogonally polarized
fundamental rogue wave governed by the coupled NLS equations with
negative coherent coupling also involves fourth-order polynomials~\cite{7}.
Additionally, a model that has been proposed in connection with spinor Bose-Einstein condensates in Ref.~\cite{xx} also involves fourth order polynomials~\cite{8}.
These studies  highlight the relevance and interest towards exploring
further the expanded palette of possibilities afforded, e.g., by such
multi-component setups.

Due to the energy transfer between different components, the central
amplitudes of the vector fundamental rogue waves are not generally
fixed,
but rather they could be varied from zero to triple that of the
background~\cite{6,7,8,tt1,tt2}.  It is relevant to also
note that the fundamental rogue waves are not the highest waves.
The highest waves may appear as a result of superpositions of
breathers, solitons and rogue waves~\cite{xj3,xj4,9}. In the case of
collisions of Peregrine solitons, the amplitude becomes larger than
that of a single Peregrine soliton~\cite{9,10}.
A recent study~\cite{10} has shown that the Peregrine soliton, still
expressed by a rational function of second degree (second-order
polynomials), can reach an amplitude limit as high as 5 times the
background level due to the energy transfer between different
components and the self-steepening effect.
Very recently, Chen et al.~\cite{special} have shown
the possibility for one
component to grow in an extreme fashion at the expense of the other
component. In particular, in this presently state-of-the-art case,
typical
results exhibit an eightfold peak amplitude increase and it is
mentioned
that a maximal enhancement of above 17 can be achieved.

In this paper, we utilize this recently emerging platform  of
vector (i.e., two-component) rogue waves but bring to bear
a drastically different mechanism. We select the rational solutions of
fourth degree as
the fundamental rogue wave solutions.
Yet, our two-component NLS variant features a crucially different,
non-sign definite mass (energy in optics)
conservation law that enables each component to grow {\it indefinitely much}
in comparison to the background.
As a result of this, we show that the vector fundamental rogue waves
(in both components) can reach a  peak amplitude as high as a {\it thousand}
times the background level due to the coherent coupling terms.
No such case occurs in integrable systems known so far, to the best of
our knowledge, and naturally it significantly eclipses the best known
current
result, as per the above discussion.
The physical relevance
of the broader class of models within which our system lies naturally
then begs the
question
of whether a physical realization of such a mechanism may be possible.
I.e., while our findings arise in a class of models that have been
used in optics and atomic physics, among other themes, we are not
aware of a physical realization of the model for the coefficient
values that our integrable model features. We note that this is often the case for
integrable models such as the so-called Ablowitz-Ladik discretization
of the NLS~\cite{AL}, and more recently the integrable spinor
NLS~\cite{xx}
or the integrable nonlocal NLS~\cite{mussli}. Nevertheless, as is the
case
with those integrable models, we expect that the present integrable
model
and its remarkable rogue wave properties will be the source of further
studies both on the physical side (to explore the realizability of
such
a setting) and on the mathematical side (to explore the model
properties, and, e.g.,
its usefulness towards perturbative treatments).
\\
\\
\noindent\textbf{\large 2. Mathematical Formulation}\\\hspace*{\parindent}

A vector NLS system with coherent and
incoherent nonlinear couplings
governing the dynamics of two orthogonally polarized modes
in a
nonlinear optical fiber is~\cite{agra}:
\begin{subequations}\label{1}
\begin{eqnarray}
&&\hspace{-0.5cm}iQ_{1z}+Q_{1tt}+2(A |Q_{1}|^2 + B |Q_{2}|^2)Q_{1}+ C Q^{*}_{1}Q^2_{2}=0,\label{m}\\
&&\hspace{-0.5cm}iQ_{2z}+Q_{2tt}+2(D |Q_{1}|^2 + E |Q_{2}|^2)Q_{2}+F Q^{*}_{2}Q^2_{1}=0,\label{n}
\end{eqnarray}
\end{subequations}
where $Q_{1}$ and $Q_{2}$ are the complex envelopes of the two field components, with $z$ and $t$ the propagation distance and retarded time, respectively.
The potential applications of FWM in coupled NLS equations have been
discussed in
numerous references; see, e.g.,~\cite{11,12,KT2013,pg}.
Typical experimentally relevant values, as discussed in~\cite{agra}
are, e.g., $A=B=D=E=1=2C=2F$. However, as discussed in,
e.g.,~\cite{12}
other combinations are physically possible, such as, e.g., $A=E=1$,
$B=D=\mu$ and $C=F=2 (1-\mu)$, where $\mu$ is a real parameter
satisfying $0<\mu<1$ (with $\mu=1/3$ being relevant, e.g.,
for dielectric materials with purely electronic response).

Here, we will use the combination of the
relevant coefficients discussed in~\cite{13} which, nevertheless, amounts to an
integrable system, namely $2A=-B=-C=D=-2E=F=2$. As indicated above,
and similarly to a number of other integrable variants, we are not
aware of a direct physical application of this setting, yet the
physical relevance of the model for different parameters renders it a
ripe testbed for mathematical and computational studies.
For this choice and if $Q_{2}\equiv k Q_{1}$ with $k$ being pure
imaginary, the model can be
reduced to the scalar NLS equation.
Furthermore, the appearance of case examples
where the inter-component interaction may feature a negative sign
in the cubic cross-coupling term (for a recent example from atomic physics, see,
e.g.,~\cite{karta}),
and on the other hand, the rather remarkable properties of the
rogue waves of this system (including the unprecedented mechanism
discussed below) suggest, in our view, the interest in considering this system
as a prototype in this vein.

Using the Darboux-dressing transformation~\cite{dt1,dt2,dt3,dt4}, we obtain the fundamental rogue wave solutions
\begin{subequations}\label{2}
\begin{eqnarray}
&&Q_{1}=\lim_{\lambda\rightarrow i \sqrt{{a_1}^2 -{a_2}^2}}a_1 e^{i \left(2 a^2_{1}-2 a^2_{2}\right) z}-2iS_{13},\label{s1}\\
&&Q_{2}=\lim_{\lambda\rightarrow i \sqrt{{a_1}^2-{a_2}^2}}i a_2 e^{i \left(2 a^2_{1}-2 a^2_{2}\right) z}-2iS_{14},\label{s2}
\end{eqnarray}
\end{subequations}
where the matrix elements are obtained from the matrices:
\begin{eqnarray}
&&S=\mathcal{A}\begin{pmatrix}  \lambda & 0&  0 & 0    \\
      0 & \lambda &  0 & 0  \\
     0 & 0 &\lambda^* &0 \\
      0 & 0& 0 & \lambda^* \end{pmatrix} \mathcal{A}^{-1},\nonumber \mathcal{A}=\begin{pmatrix} \psi_{1} &-\psi_{2}&  \psi^*_{3} &-\psi^*_{4}    \\
      \psi_{2} & \psi_{1} &  \psi^*_{4} & \psi^*_{3}  \\
     \psi_{3} & -\psi_{4} & -\psi^*_{1} & \psi^*_{2} \\
     \psi_{4} &\psi_{3}& -\psi^*_{2} &-\psi^*_{1} \end{pmatrix},
\end{eqnarray}
\begin{eqnarray}
                                       \begin{pmatrix} \psi_{1}\\ \psi_{2}\\ \psi_{3}\\ \psi_{4} \end{pmatrix}=\mathcal{F} e^{\mathcal{B}(\lambda) z +\mathcal{C}(\lambda) t} \mathcal{F}_{0},
        \end{eqnarray}
with
\begin{eqnarray}
&&\mathcal{F}=\left(
\begin{array}{cccc}
 1 & 0 & 0 & 0 \\
 0 & 1 & 0 & 0 \\
 0 & 0 & e^{-i z \left(2 {a_1}^2-2 {a_2}^2 \right)} & 0 \\
 0 & 0 & 0 & e^{-i z \left(2 {a_1}^2-2 {a_2}^2 \right)} \\
\end{array}
  \right),
\end{eqnarray}
\begin{eqnarray}
  \mathcal{C}(\lambda)=\left(
\begin{array}{cccc}
 -i \lambda  & 0 & a_1 & i a_2 \\
 0 & -i \lambda  & -i a_2 & a_1 \\
 -a_1 & i a_2 & i \lambda  & 0 \\
 -i a_2 & -a_1 & 0 & i \lambda  \\
\end{array}
\right),\nonumber
\end{eqnarray}

\begin{equation}
\mathcal{B}(\lambda)=\left(
\begin{array}{cccc}
 -i \left(2 \lambda ^2-a_1^2+a_2^2\right) & 0 & 2 \lambda  a_1 & 2 i \lambda  a_2 \\
 0 & -i \left(2 \lambda ^2-a_1^2+a_2^2\right) & -2 i \lambda  a_2 & 2 \lambda  a_1 \\
 -2 \lambda  a_1 & 2 i \lambda  a_2 & i \left(2 \lambda ^2+a_1^2-a_2^2\right) & 0 \\
 -2 i \lambda  a_2 & -2 \lambda  a_1 & 0 & i \left(2 \lambda ^2+a_1^2-a_2^2\right) \\
\end{array}
\right).\nonumber
\end{equation}
Here, $\mathcal{F}_{0}=(z_{1},z_{2}, z_{3}, z_{4})^{T}$ is a complex constant vector, $a_{1}$ and $a_{2}$ are real constants corresponding to the background heights, $S_{1j}$ $(j=3,4)$ represents the entry of matrix $S$ in the $1$st row and $j$th column.
We have provided the analytical forms of rogue wave solutions used
in the Appendix. In general, such fundamental
rogue-wave solutions are expressed by rational functions of fourth
degree, but not by ones of second degree. When $Q_{2}\equiv
\frac{a_2}{a_1}i Q_{1}\ (\left|\frac{a_2}{a1}\right|<1)$, such
fundamental rogue-wave solutions revert to the rational functions of
second degree, which are the same as the Peregrine soliton of the focusing NLS equation.

It is relevant to highlight that the conserved ``energy'' (in the context of
optics) for the model of Eq.~(\ref{m})-(\ref{n}) reads:
\begin{eqnarray}
E=\int_{-\infty}^{\infty}|Q_{1}|^2-|Q_{2}|^2 dt,
\label{ener}
\end{eqnarray}
and differs significantly from the customary conservation of the sum
of the two-component squared $L^2$ norms in FWM models~\cite{agra} (or their individual
conservation, e.g., in Manakov-type models~\cite{agra,AL}).
This is particularly crucial because it enables the dynamical
evolution of the mass
of the two components in a way such that both may grow {\it
  indefinitely} with respect to the background,
but retain their relative size with respect to each other so that the
conservation
law of Eq.~(\ref{ener}) is preserved. This appears to be the principal
mechanism
at work, enabling the dramatic intensity enhancements that will arise
parametrically in the examples that will follow in the numerical
illustration below. For completeness, additional conservation laws
of the integrable model at hand, such as its Hamiltonian ($\mathcal{H}$) and momentum ($\mathcal{M}$)
are given as~\cite{13}:
\begin{equation}
\begin{array}{c}
\mathcal{M}=i \int_{-\infty}^{+\infty}\left(Q_{1} Q_{1 t}^{*}-Q_{1}^{*} Q_{1 t}-Q_{2} Q_{2 t}^{*}+Q_{2}^{*} Q_{2 t}\right) \mathrm{d} t, \\
\mathcal{H}=\int_{-\infty}^{+\infty}\left[\left(\left|Q_{1 t}\right|^{2}+\left|Q_{2 t}\right|^{2}\right)-\left(\left|Q_{1}\right|^{4}+\left|Q_{2}\right|^{4}\right)\right. \\
\left.\quad+4\left|Q_{1}\right|^{2}\left|Q_{2}\right|^{2}+\left(Q_{1}^{* 2} Q_{2}^{2}+Q_{1}^{2} Q_{2}^{* 2}\right)\right] \mathrm{d} t.
\end{array}
\end{equation}

\noindent\textbf{\large 3. Numerical Verification}\\\hspace*{\parindent}

\begin{figure}
\hspace{-1cm}\includegraphics[scale=0.400]{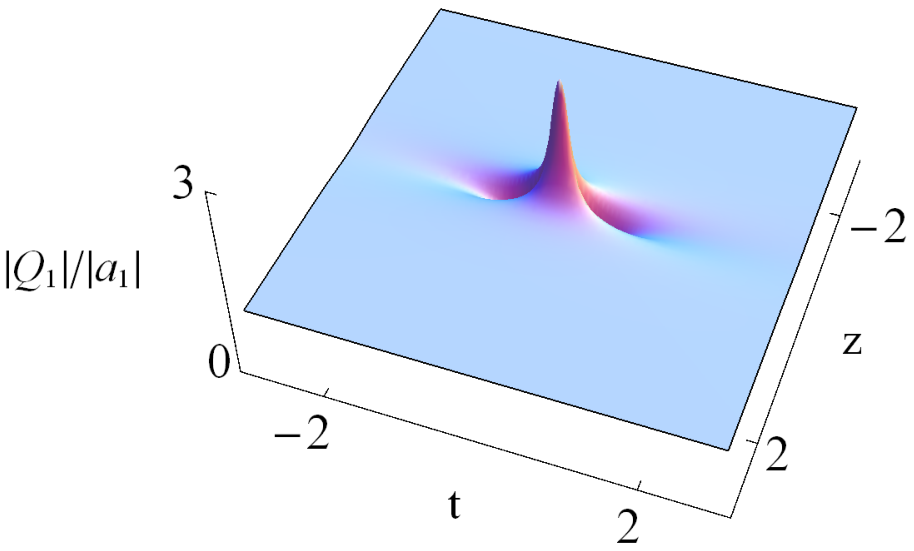}\quad \includegraphics[scale=0.400]{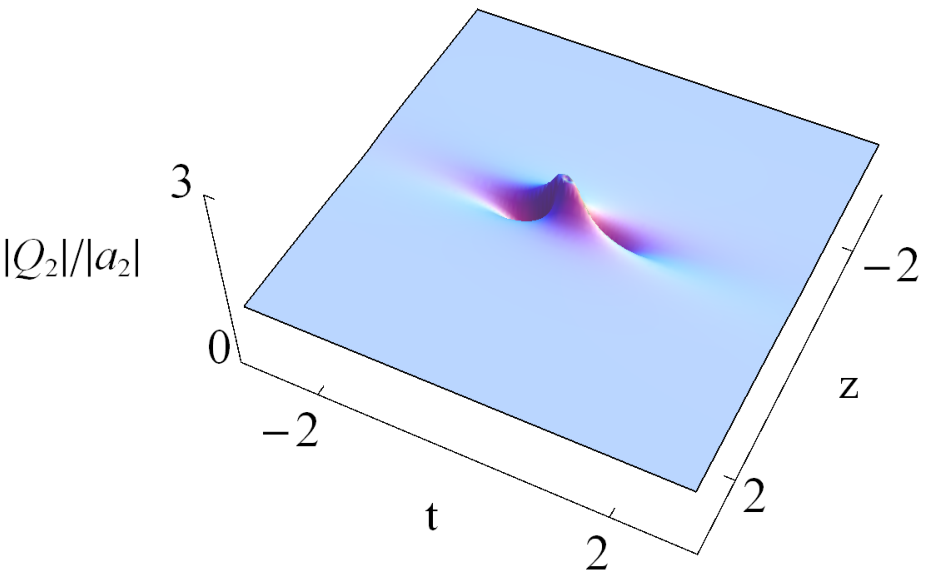}\\
\hspace{-0.7cm}\includegraphics[scale=0.400]{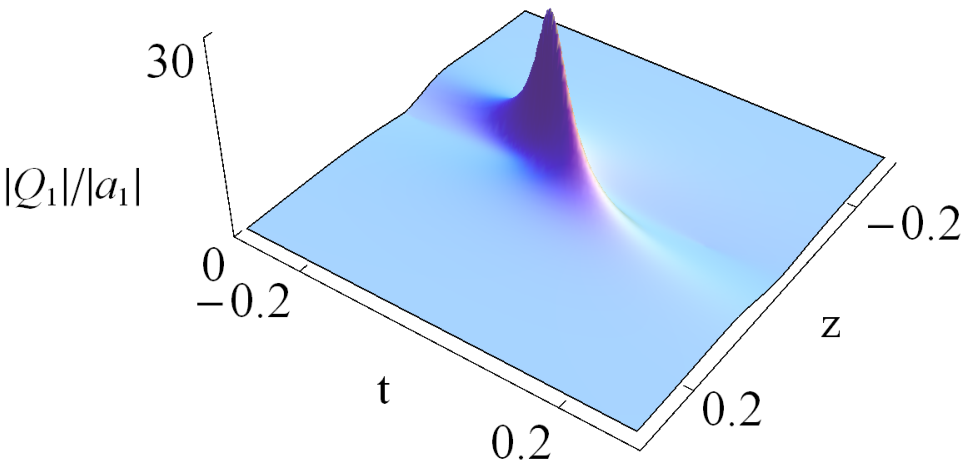}\quad \includegraphics[scale=0.400]{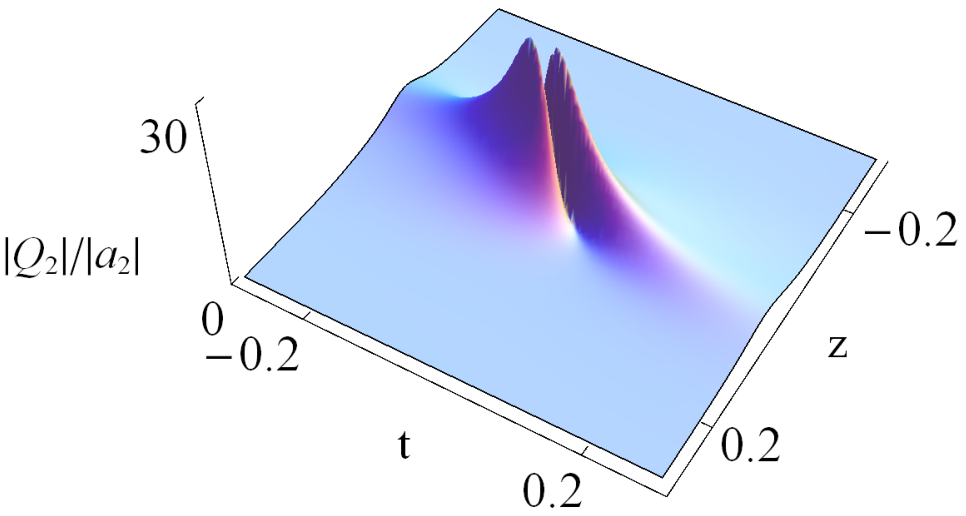}\\
\hspace{-0.7cm}\includegraphics[scale=0.400]{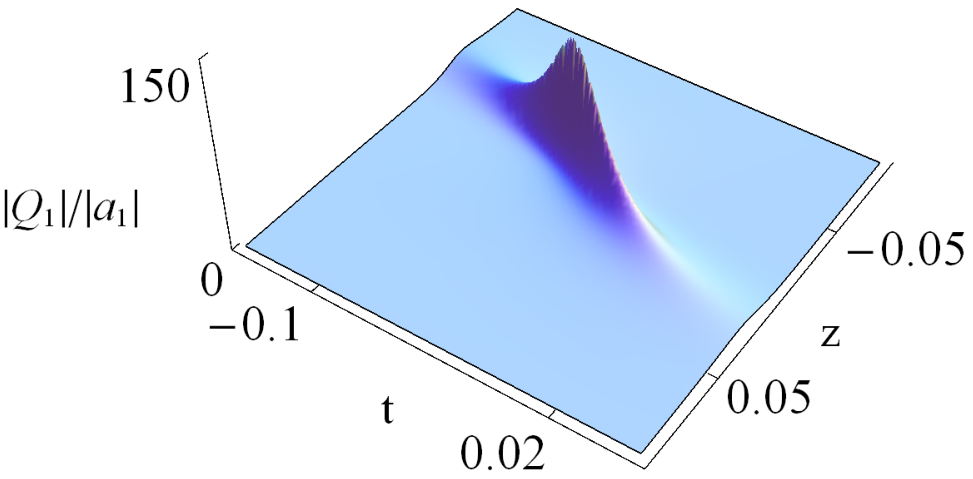}\quad \includegraphics[scale=0.400]{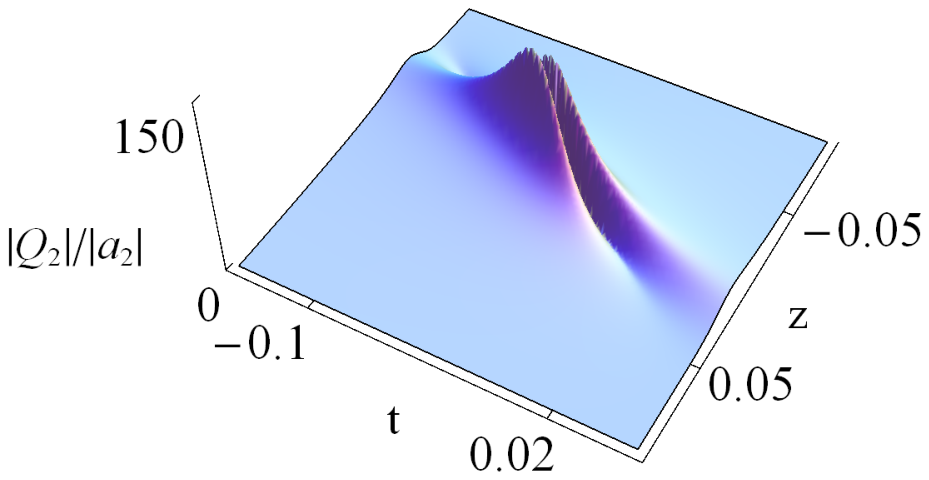}\\
\hspace{-0.7cm}\includegraphics[scale=0.400]{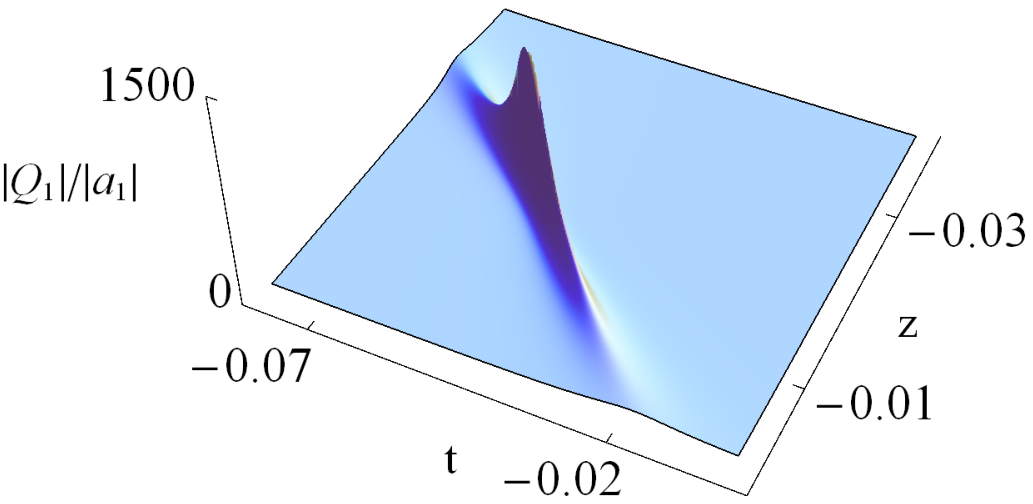}\quad \includegraphics[scale=0.400]{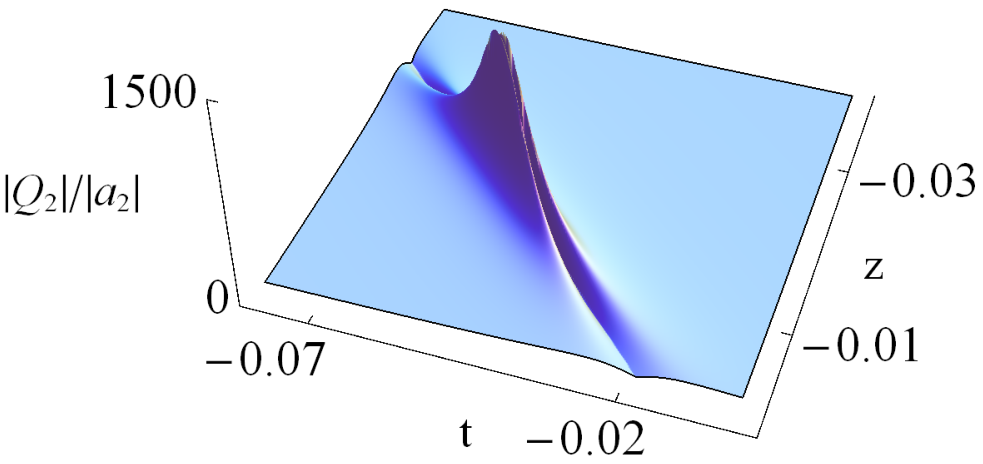}\\
\caption{\label{fig:epsart1}
  The top panel shows the fundamental
  rogue wave dynamics of our solutions~(\ref{2}),  obtained with
the parameters $a_{1}=2$, $a_{2}=1$, $z_{1}=3+i$, $z_{3}=0$, $z_{4}=3$ and $z_{2}=4$.
  The same is shown for different parameters  $z_{2}=1/3$ (2nd row), $z_{2}=1/6$ (3rd row) and $z_{2}=12/100$ (4th row).}
\end{figure}

To illustrate the fundamental rogue-wave
dynamics, we use the parameters $a_{1}=2$, $a_{2}=1$, $z_{1}=3+i$,
$z_{3}=0$, $z_{4}=3$, but with different structural parameter $z_{2}$
in the panels of Fig.~\ref{fig:epsart1}.
In the top panel, we show the standard fundamental rogue-wave dynamics of our solutions~(\ref{2}),  obtained with
structural parameter $z_{2}=4$. It is seen that the vector fundamental rogue waves reach the amplitude as high as
$0\sim3$ times the background level. Such a realization can be viewed
as the standard case, which can also be observed in the SS
equation~\cite{6,tt1}, as well as in the coupled NLS equations of Refs.~\cite{7,8}.
However, in stark contrast with the previous
studies~\cite{6,7,8,tt1,dt2}, we will
show that the fundamental rogue waves of Eqs.~(\ref{1}) could reach
an extremely
high peak amplitude.

\begin{figure}
\hspace{-0.4cm}\includegraphics[scale=0.3100]{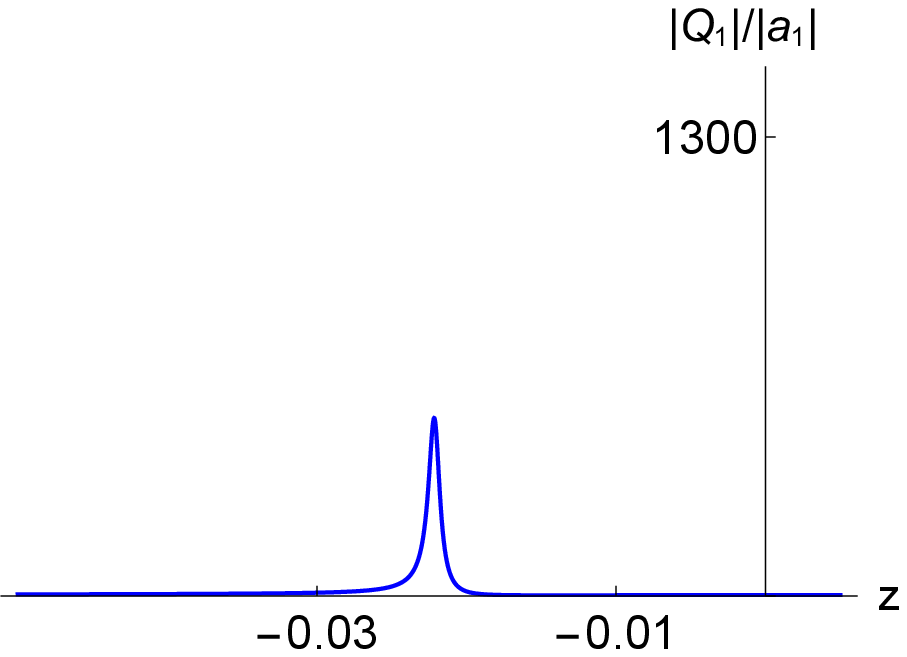}\hspace{0.15cm}\includegraphics[scale=0.3100]{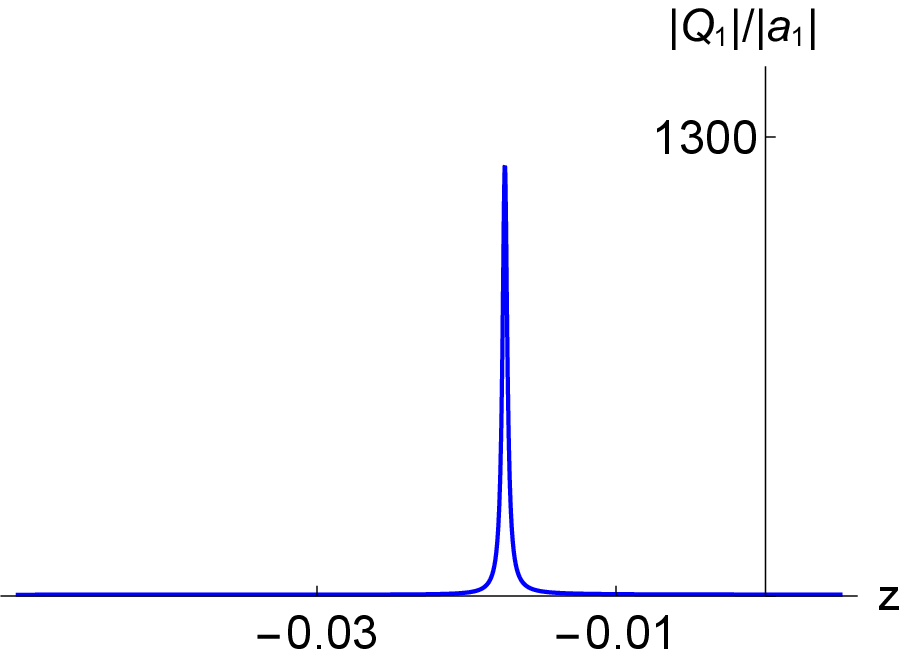}\hspace{0.15cm} \includegraphics[scale=0.31000]{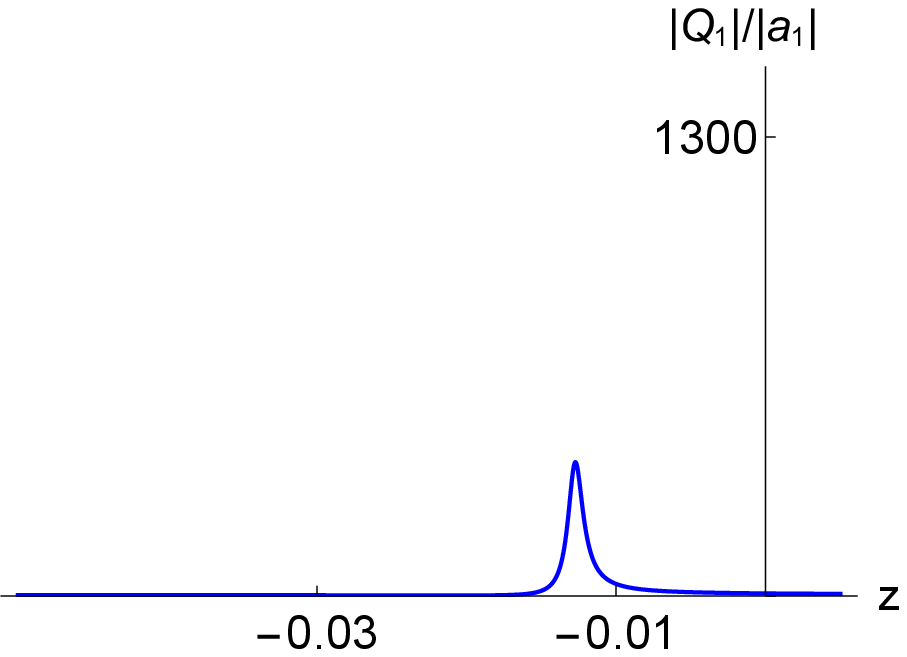}\\
\hspace{-0.4cm}\includegraphics[scale=0.3100]{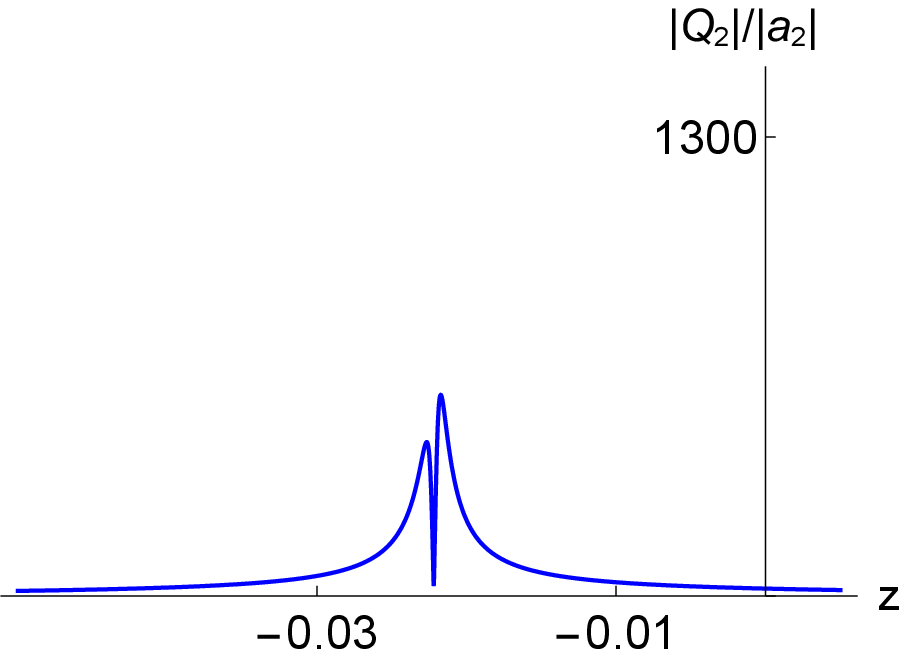}\hspace{0.15cm}\includegraphics[scale=0.3100]{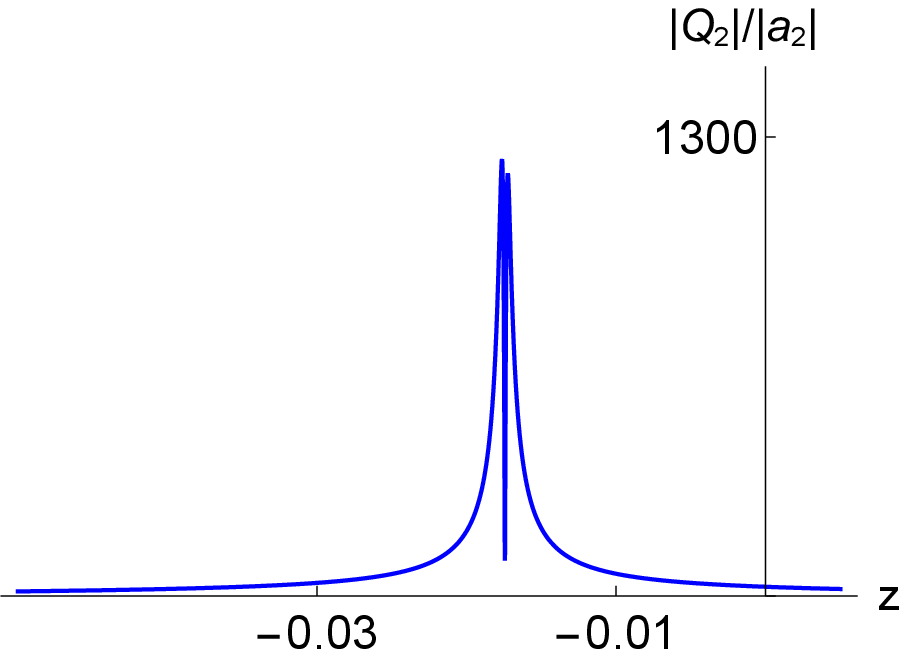}\hspace{0.15cm} \includegraphics[scale=0.3100]{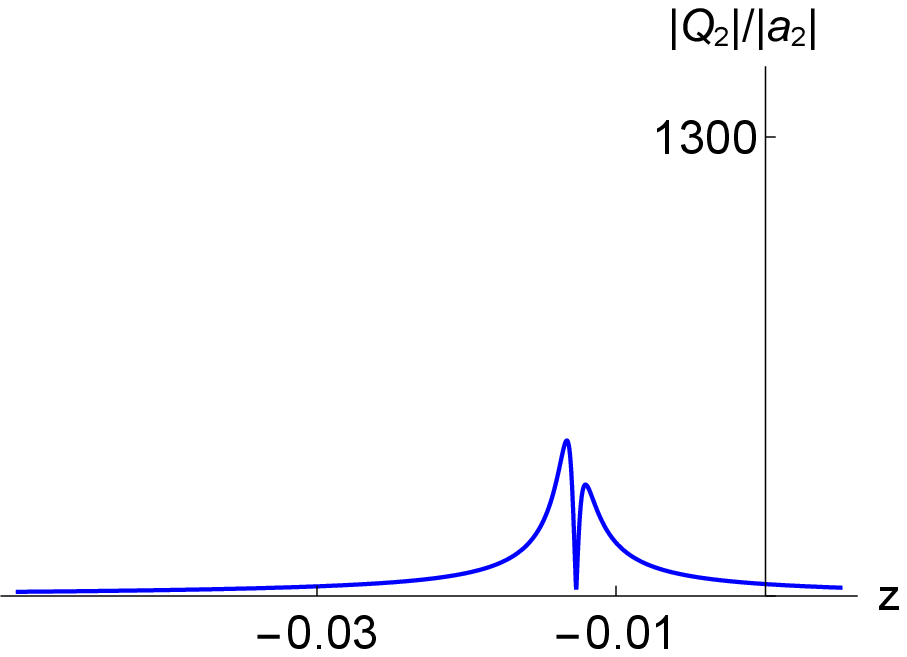}
\caption{\label{fig:epsart2} The fundamental
  rogue wave dynamics of our solutions,  obtained with
the parameters $a_{1}=2$, $a_{2}=1$, $z_{1}=3+i$, $z_{3}=0$,
$z_{4}=3$, $z_{2}=12/100$ and $t= -0.06, -0.05, -0.04$ (from left to
right). The modulus of each field is shown (top row: $Q_1$ and bottom
row $Q_2$) normalized to its respective background amplitude.}
\end{figure}

\begin{figure}
\includegraphics[scale=0.3400]{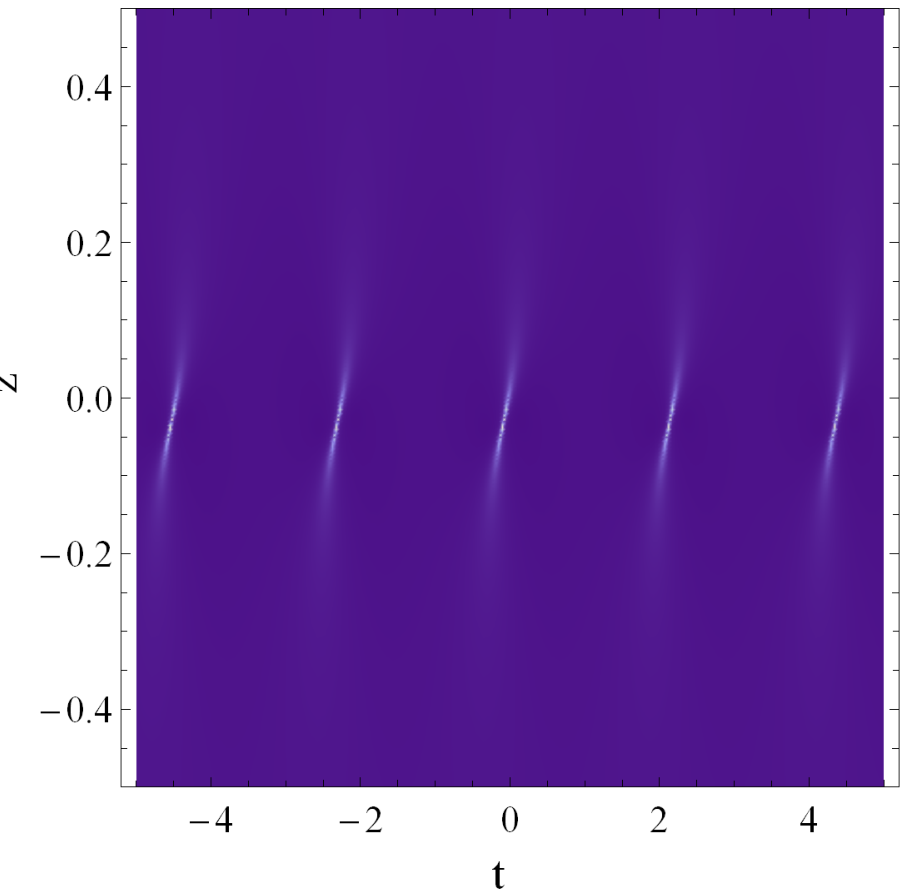}\ \includegraphics[scale=0.5400]{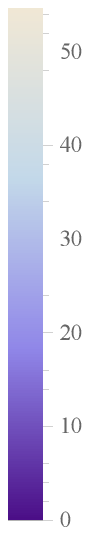}\ \ \includegraphics[scale=0.3400]{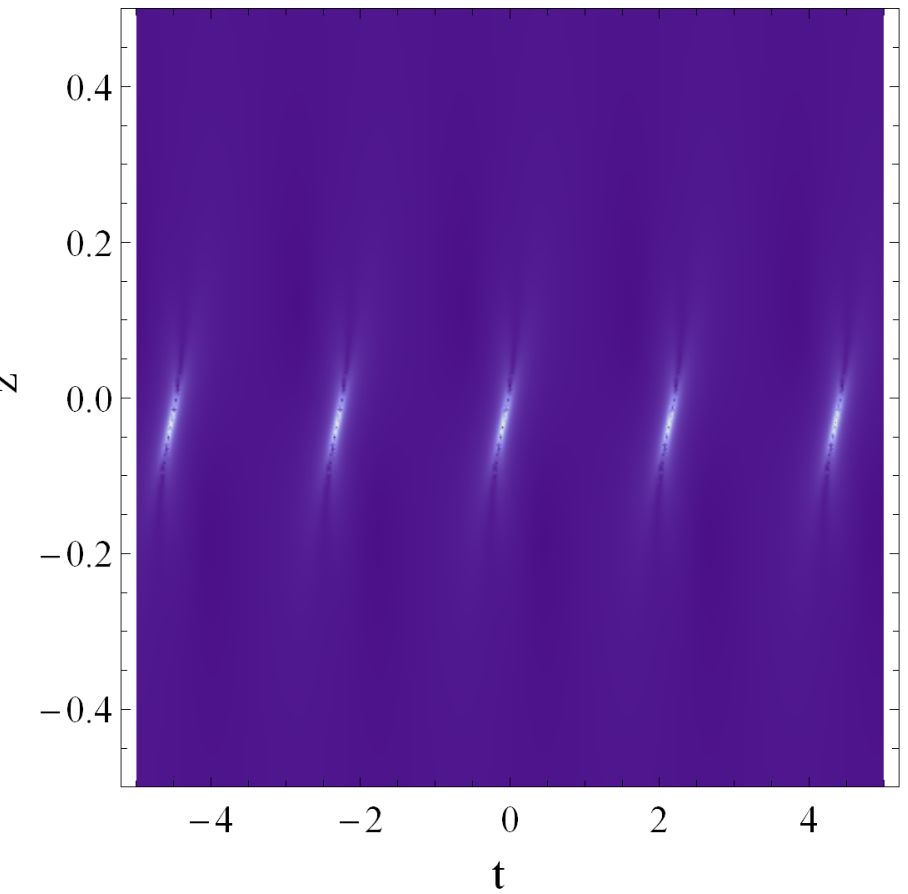} \ \includegraphics[scale=0.5400]{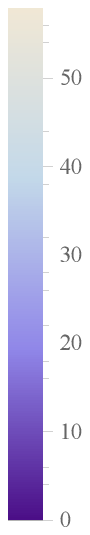}\\
\includegraphics[scale=0.3400]{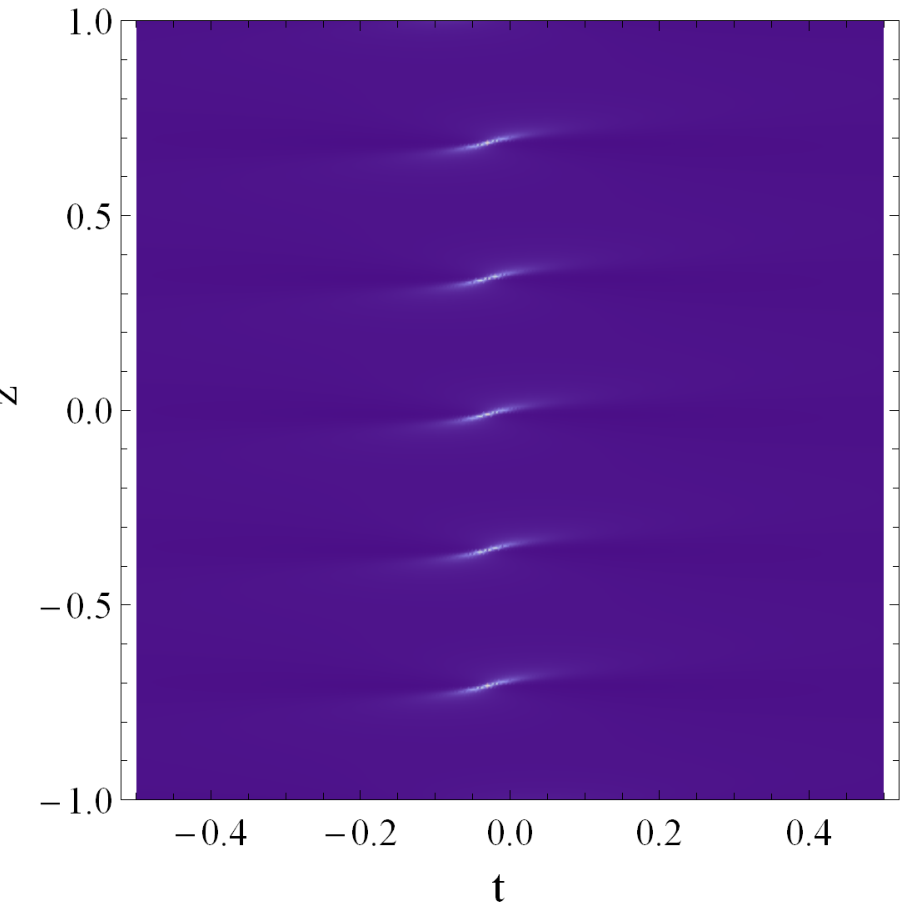}\ \includegraphics[scale=0.5400]{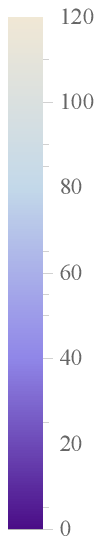}\ \ \includegraphics[scale=0.3400]{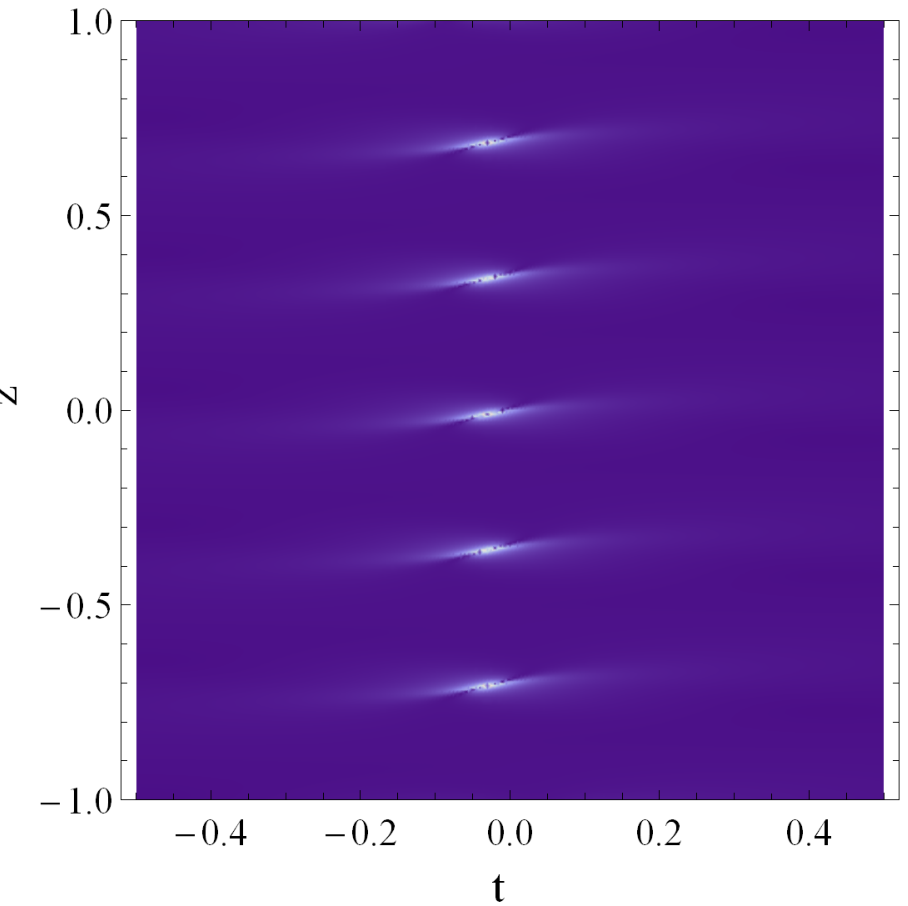} \ \includegraphics[scale=0.5400]{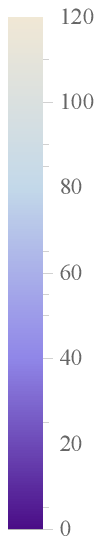}\\
\caption{\label{fig:epsart3} Space-time contour plots of
  vector breathers with $z_{1}=3+i$,
  $z_{2}=1/6$, $z_{3}=0$, $z_{4}=3$, $a_{1}=2$, $a_{2}=1$,
  $\lambda=i<\sqrt{3}i$ (Akhmediev breather in the top row) and
  $\lambda=2.5i>\sqrt{3}i$ (Kuznetsov-Ma soliton in the bottom row).}
\end{figure}

In the remaining panels of the figure, depending on the relative values of the structural parameter $z_{2}$,  the fundamental rogue wave
solutions, still expressed by  rational functions of the fourth
degree, can remarkably reach an amplitude  as high as a
thousand times the background level. When $z_{2}=\frac{1}{3}$, it is
noted that
$Q_{1}$ and $Q_{2}$ reach an amplitude
limit at least  as high as
$26$ times the background level. When $z_{2}=\frac{1}{6}$, it is noted
that
both components are at least over
$110$ times the background level. When $z_{2}=\frac{12}{100}$, it is
noted that
$Q_{1}$ and $Q_{2}$ reach an amplitude limit as high as at least
$1200$ times the background level. Besides, the $Q_{1}$ component has one
peak,  while the $Q_{2}$ component has two peaks. This is in line with the
feature that such equations also admit the vector single
peak-double peak solitons on top of a vanishing background~\cite{13}.
For completeness, in Fig.~\ref{fig:epsart2}, we illustrate the
profile of the
modulus of both spatial fields (divided by their respective
backgrounds) near and at the instance of peak formation.
We show how the fundamental rogue waves reach
the ultra-high peak amplitude in both fields within a short
time. It is worth noticing how the second component forms a peak first on one side
of the first component maximum, subsequently symmetrizes and then
the peak appears on the other side of the first component maximum.
It is also relevant to note that
when $|\lambda|<\sqrt{{a_1}^2 -{a_2}^2}$,  Akhmediev breathers arise
in the model, which means that the solutions exhibit
localization in $z$ but periodicity along $t$. On the other hand,
when $|\lambda|>\sqrt{{a_1}^2 -{a_2}^2}$,  Kuznetsov-Ma solitons
appear, which means that the solutions exhibit localization in $t$ but
periodicity along $z$. Akhmediev breathers and Kuznetsov-Ma solitons
with ultra-high peak amplitude are shown in
Fig.~\ref{fig:epsart3}.

\begin{figure}
\hspace{-1cm}\includegraphics[scale=0.5500]{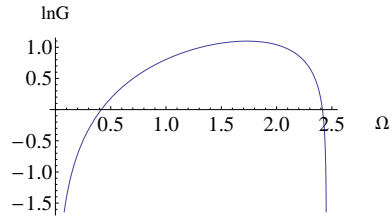}
\caption{\label{fig:epsart4}  MI gain at  $k=0, a_{1}=2, a_{2}=1$.}
\end{figure}

Now we show that such vector rogue waves could be generated in
the  modulation instability (MI) regime.
We take the background solutions as $Q_{10}=a_{1}e^{i (k z+ \omega t)}$
and $Q_{20}=ia_{2}e^{i (k z+\omega  t)}$ with $k=2 a_1^2-2
a_2^2-\omega^2$.
A perturbed nonlinear
background can be written as $Q_{1}=(a_{1}+p_{1})e^{i (k z+\omega t)}$
and $Q_{1}=i(a_{2}+p_{2})e^{i (k z+\omega  t)}$, where $p_{1}$ and
$p_{2}$ are small
perturbations. The
$p_{1}$ and $p_{2}$ are  $t$-periodic with frequency $\Omega$. Using linear stability analysis, we obtain the eigenvalues of the linear system as $\pm \sqrt{-4 a_1^2 \Omega ^2+4 a_2^2 \Omega ^2+\Omega ^4}-2 \omega \Omega$.
When the eigenvalue has a negative imaginary part, MI arises, which means $a_{1}>a_{2}$. Therefore, when $|a_{1}|>|a_{2}|$,  a baseband MI~\cite{pr4}, which
includes frequencies that are arbitrarily close to zero, is
present, i.e., $0<\Omega^2<4a^2_{1}-4a^2_{2}$. In Fig.~\ref{fig:epsart4},
we show the logarithmic gain plot $\ln G({\Omega})$ versus $\Omega$, where $G(\Omega)=\sqrt{+4 a_1^2 \Omega ^2-4 a_2^2 \Omega ^2-\Omega ^4}$.
The gain maximum is found to be $1.09861$ at $\Omega=1.73205$.

\begin{figure}
\hspace{-0.4cm}\includegraphics[scale=0.3600]{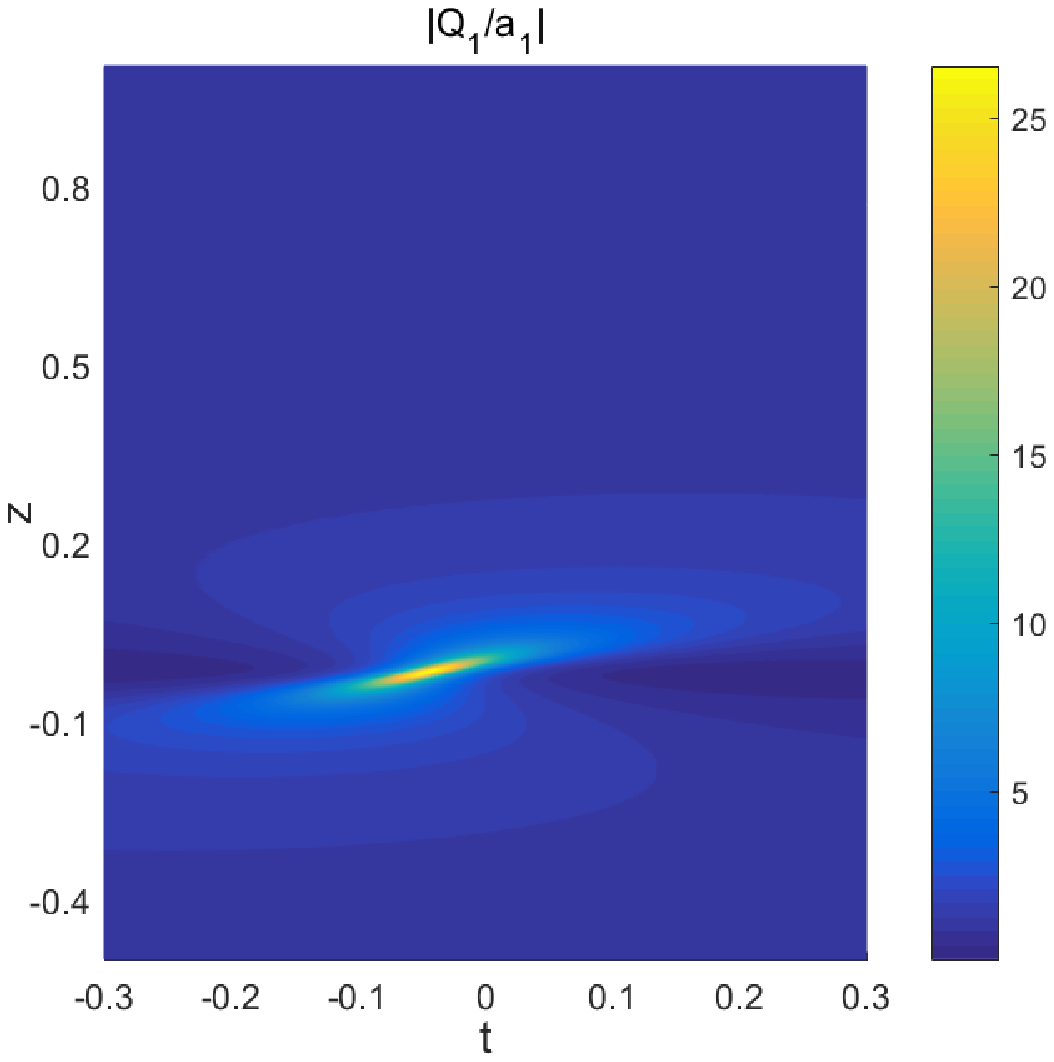}\hspace{0.15cm} \includegraphics[scale=0.3600]{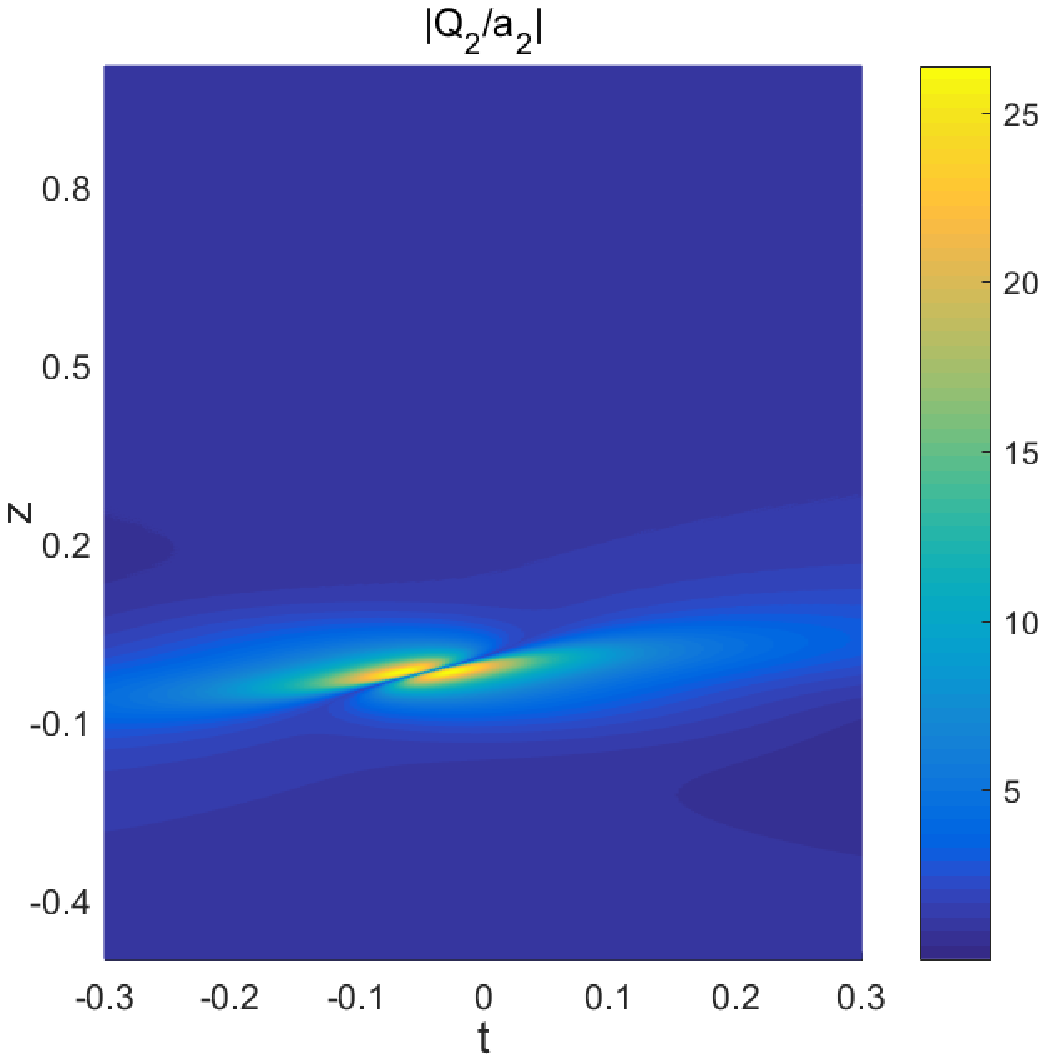}\\
\hspace{-0.4cm}\includegraphics[scale=0.3600]{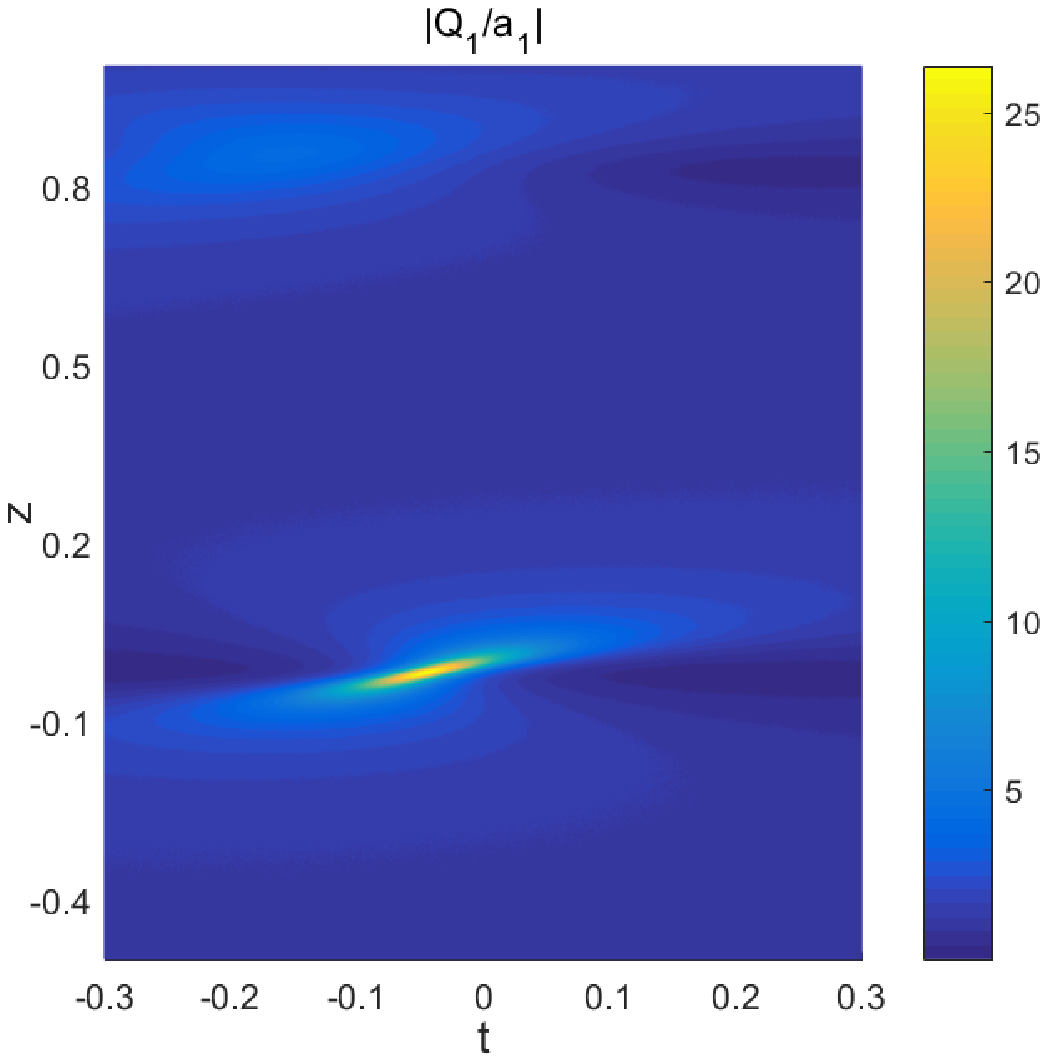}\hspace{0.15cm} \includegraphics[scale=0.3600]{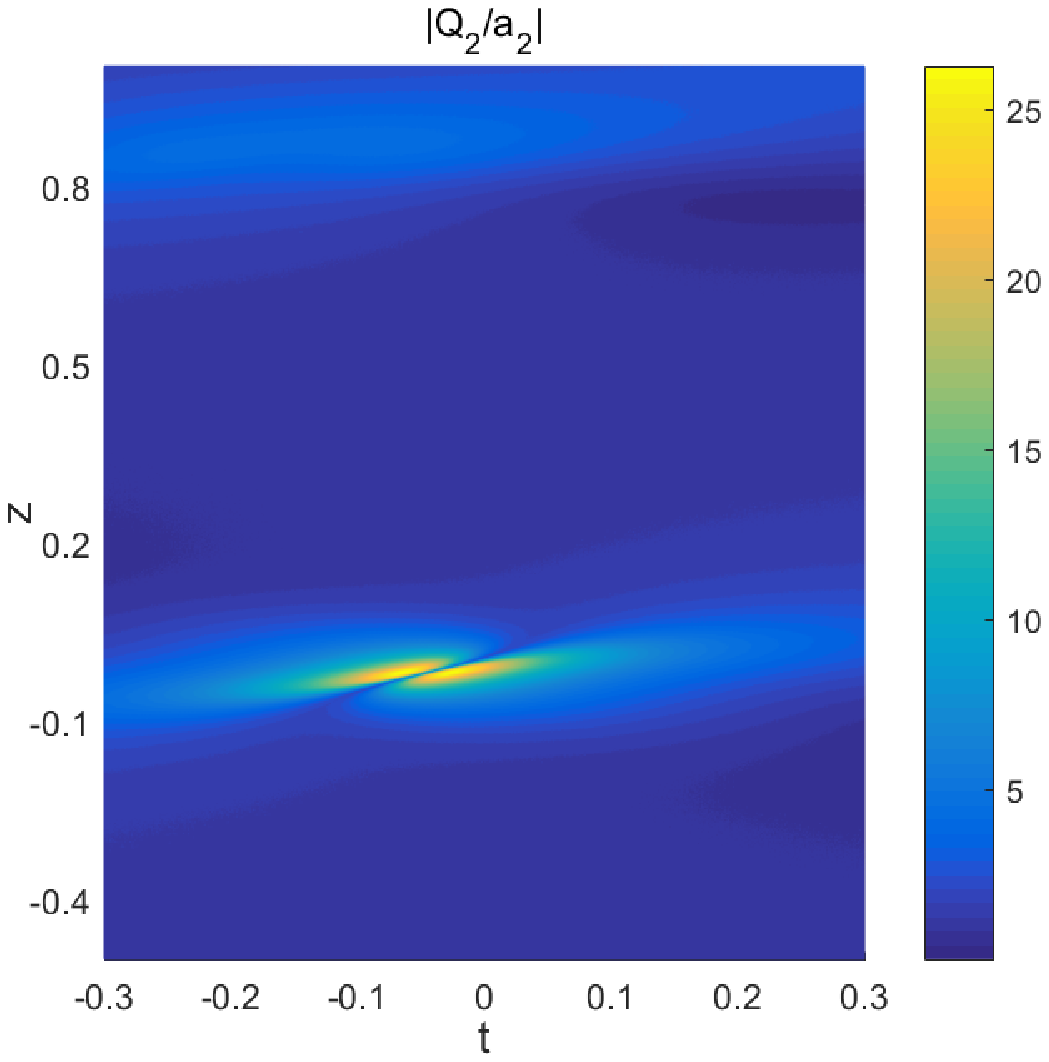}
\caption{\label{fig:epsart5} Numerical simulations of the $z-t$
  evolution (shown through contour plots) of the fundamental rogue waves in
  Fig.~\ref{fig:epsart1}
  with $z_{2}=1/3$  without perturbation (upper row) and with $5\%$ white noise perturbation (bottom row).}
\end{figure}


We also use numerical simulations to investigate the robustness of
these fundamental rogue waves in Fig.~\ref{fig:epsart5}. Here, the
split-step Fourier spectral method is used to deal with the spatial
derivative operators and the fourth-order Runge-Kutta method is
brought to bear
to tackle the forward time marching of Eqs.~(\ref{1}).
We can observe that both
with (bottom row panels) and  without (top row panels)
imposing a $5\%$ white noise perturbation~\cite{tt1}
on top of the solutions with $z_2=1/3$ (2nd row of
Fig.~\ref{fig:epsart1}), the waveforms are found to robustly persist in
the evolution dynamics. In the presence of a perturbation, due
to the spontaneous MI of the homogeneous state, both fields develop an
unstable background after the fundamental rogue wave propagation.
Similar case scenarios have been explored for
other values of the parameters (e.g. for the case of $z_2=1/6$ of the
third row of Fig.~\ref{fig:epsart1}) and the results of
Fig.~\ref{fig:epsart5}
have been found to be representative of the dynamical robustness of
the states at hand.

Importantly, we even find that such unusual extremely high peak-amplitude
fundamental rogue waves can be excited in a chaotic background field.
To do this, we use the plane wave solutions as initial conditions at $z=0$, perturbed by random noise of a strength of $2\%$. Specifically, we multiply the plane waves in $Q_1$ and $Q_2$ by the factors $[1+0.02f_j(t)](j=1,2)$, respectively, where $f_j(t)$ are real random functions whose mean value is $0$ and variance is $1/3$. Besides, $f_j(t)$ are taken as Gaussian distributed and Gaussian correlated functions with the correlation length $1/2$~\cite{NJA2009}.

\begin{figure}
\hspace{-0.4cm}\includegraphics[scale=0.41]{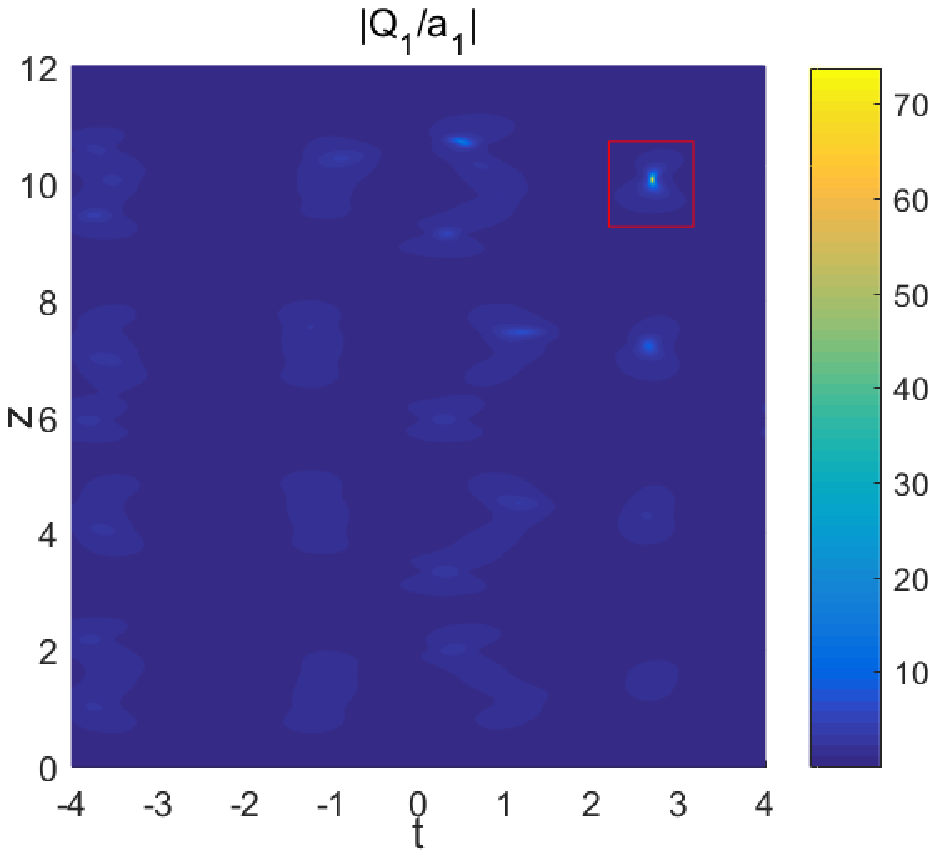} \includegraphics[scale=0.405]{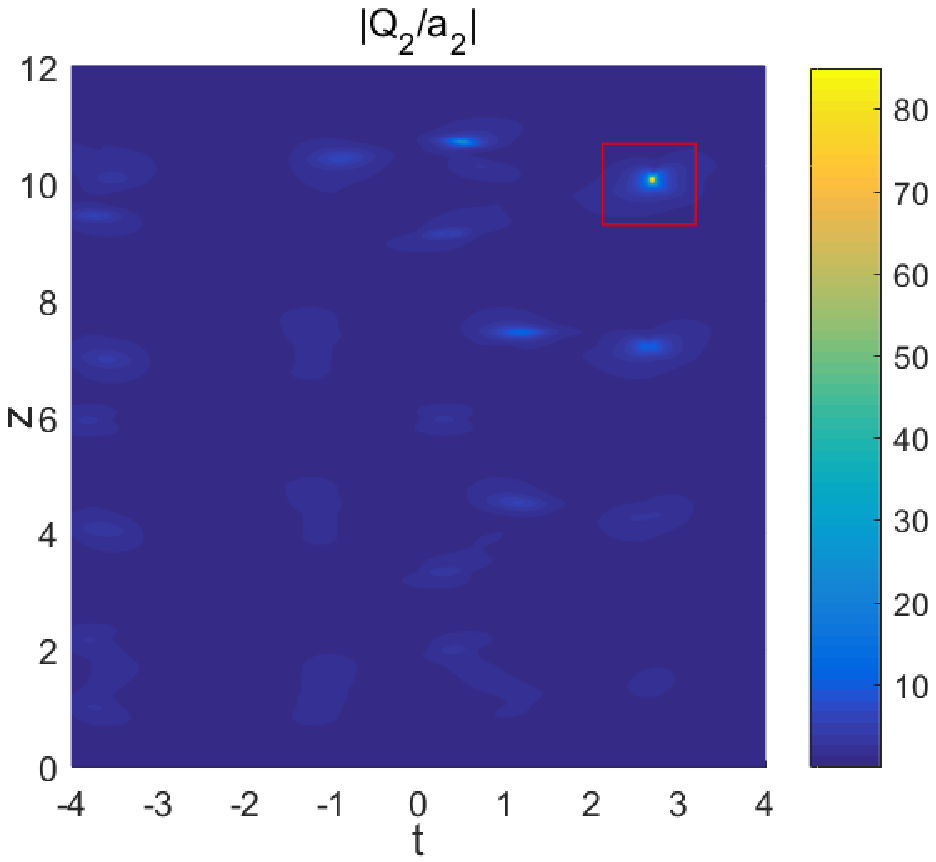}\\
\hspace{-0.4cm}\includegraphics[scale=0.355]{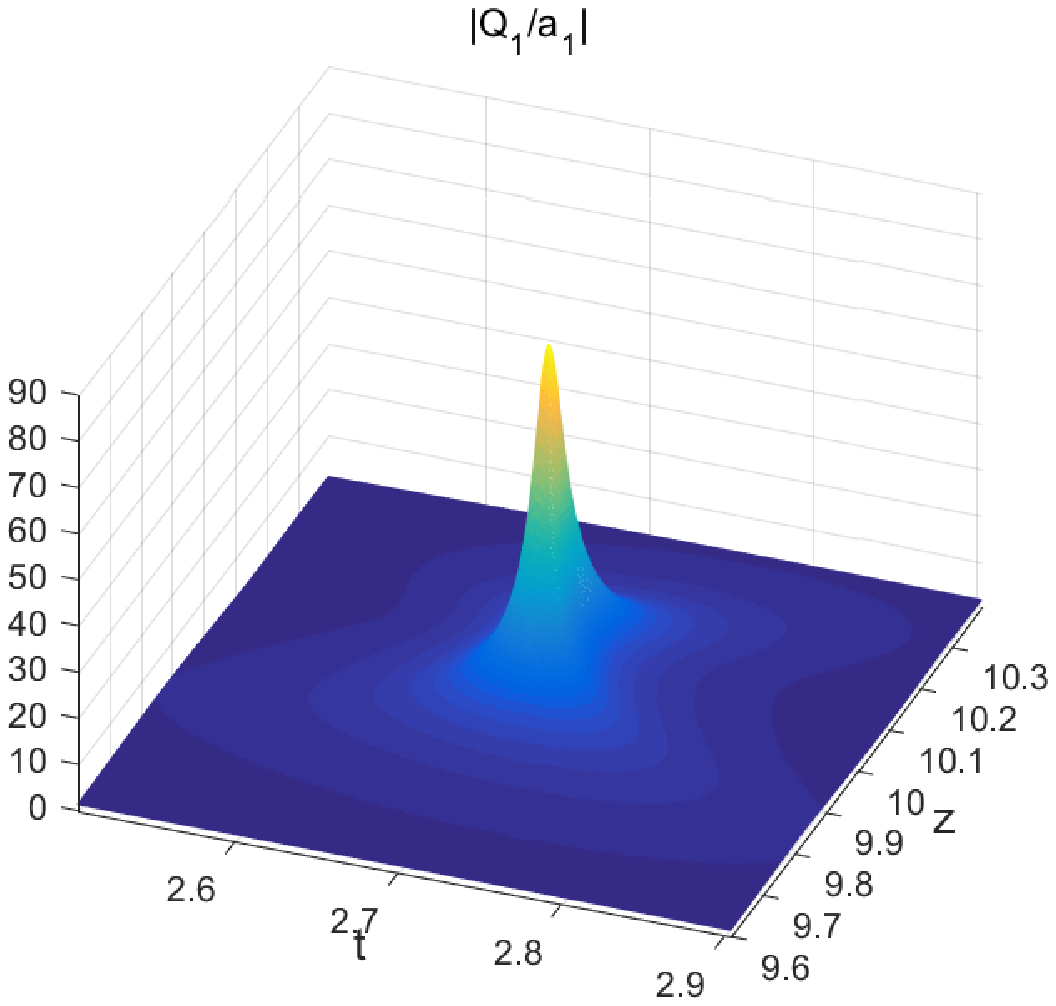} \includegraphics[scale=0.355]{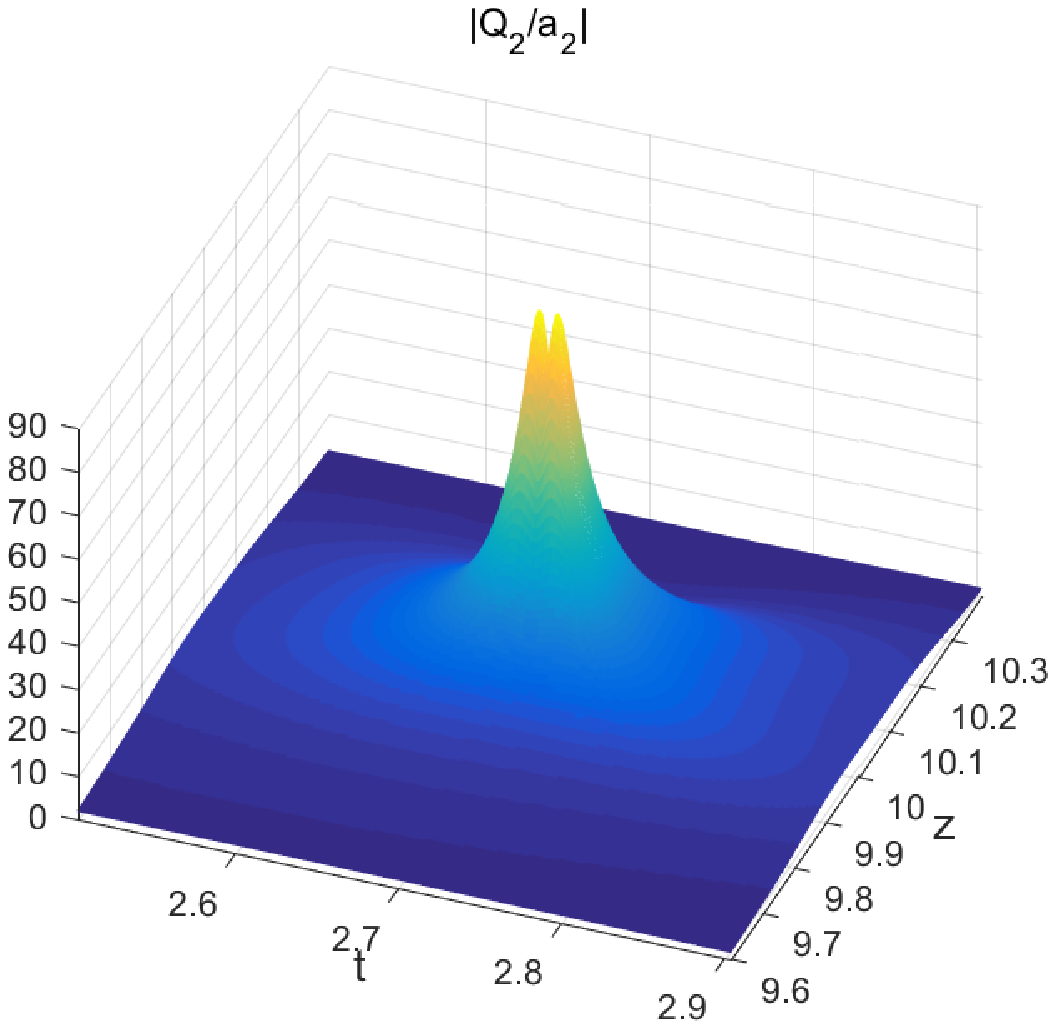}
\caption{\label{fig:epsart6} Numerical excitation of the fundamental rogue waves. The initial condition is a plane wave perturbed by $2\%$ random noise with $a_{1}=2$, $a_{2}=1$ and $f_1(t)\neq f_2(t)$: The amplitude evolution (upper row); The enlarged 3D plots of the high peak-amplitude vector rogue wave profiles highlighted by a surrounding box (bottom row).}
\end{figure}


We have examined a series of cases where  $f_1(t)=f_2(t)$. In that
case, the effective correlation built between the fields does not
allow the maximal amplitude to be significantly larger than $3$.
I.e., we are in an effective single-component NLS regime and while
rogue
waves are clearly discernible, they are not of the anomalously large
amplitude variety created by the mechanism proposed herein.
However, in Fig.~\ref{fig:epsart6}, considering the same selection of
model parameters as before, we select $f_1(t)$ and $f_2(t)$ to be
different random
variables
(empirically, we find that
even a small difference between the two is sufficient).
Then, we see that some waves in the chaotic background clearly feature
significant increase in their amplitudes. The parts selected by red
boxes are
enlarged in the corresponding mesh plots for the two fields of the
bottom
panel, revealing a high peak-amplitude vector rogue wave excited.
Such a rogue wave is
clearly distinguished from general higher-order rogue waves
(superposition of fundamental rogue waves), arising in the form of
suitable
analytical solutions, e.g., in the
realm
of the single-component NLS model~(\ref{2}). Hence, this spontaneously
emerging pattern within the chaotic background is a definitive manifestation
of the type of wave pattern advocated in the present work.
We attribute the key differences to the coherent coupling effect. Due to the existence of
coherent coupling terms in Eqs.~(\ref{1}), the phase-dependent
contribution (coherence) of plane waves in $Q_1$ and $Q_2$  plays an
important role. Indeed, that coherent coupling is chiefly responsible
for
the presence of a single energy/mass-type conservation law as per
Eq.~(\ref{ener}).
If $C=F=0$, then the two components individually conserve their
energy. In that case,
this type of two-component interplay of mutual growth leading in pairwise
cancellation at the level of Eq.~(\ref{ener}), yet indefinite growth
at the
level of each individual component is prohibited by the individual
conservation
laws. We thus argue that this mechanism and phenomenology is
{\it unique} to this type of coherently coupled NLS models and
thus not encountered in the settings previously considered.
\\
\\
\noindent\textbf{\large 4. Conclusions/Future Work}\\\hspace*{\parindent}

In the present work, we have unveiled
an
unprecedented mechanism, to the best of our knowledge, regarding the
formation of arbitrary amplitude rogue waves.
Within this scenario, we have explained the role of coherent coupling terms,
in conjunction with an unconventional (single) energy conservation law
bearing indefinite sign between the energies of individual components.
As a result, in a prototypical (within this class of features) NLS model inspired by
the examination of multiple polarizations in a nonlinear fiber,
we have shown that vector fundamental rogue waves (in both components)
of such coupled equations could reach peak amplitudes of the order of
a thousand times the background level. The fundamental solutions we
consider here
are rational solutions of fourth degree. Their periodic extensions in
the $z$ (Kuznetsov-Ma soliton) and $t$ (Akhmediev breather) variables
were also identified and illustrated.
Furthermore, we have numerically confirmed that such unusual, extremely
high peak-amplitude vector fundamental rogue waves are robust even
under
moderate perturbations. Last but not least, we have confirmed that
such vector fundamental rogue waves
could be
excited in a
chaotic background field, i.e., that such excitations indeed
spontaneously
arise in dynamical simulations against the backdrop of an unstable background.
In as far as we can tell, the key mechanism elucidated here
enables the significant
eclipsing of the highest amplitude of previously reported rogue waves.
This most naturally poses the question of whether a direct (or an
engineered)
observation of this phenomenon can arise and can be accordingly
harnessed.
From a theoretical perspective numerous extensions of the ideas
reported herein can also be pursued including the extension and
numerical
identification of such states beyond the integrable limit considered
herein, utilizing, e.g., recent ideas such as those of~\cite{cory}.
Another extension of interest is to consider two-dimensional
variants of the present model and whether rogue wave of the present
type can be identified in suitable generalizations of settings
related to the Davey-Stewartson and Benney-Roskes models where two-dimensional
rogue patterns were recently considered in~\cite{jingsong}.

\vspace{2mm}
\noindent\textbf{\large Acknowledgments}\\\hspace*{\parindent}
 This work has been supported by the National Natural Science
Foundation of China under Grant Nos.61705006, 11305060 and 11947230,
by the Fundamental Research Funds of the Central Universities
(Nos.230201606500048 and 2018MS048), and by the China Postdoctoral
Science Foundation (No. 2019M660430).
This material is based upon work supported by the US National Science
Foundation under Grants No. PHY-1602994 and DMS-1809074
(PGK). PGK also acknowledges support from the Leverhulme Trust via a
Visiting Fellowship and thanks the Mathematical Institute of the University
of  Oxford for its hospitality during part of this work.

*liueli@126.com

\noindent\textbf{\large Appendix:}\\\hspace*{\parindent} The fundamental
solutions~(\ref{2}) with
the parameters $a_{1}=2$, $a_{2}=1$, $z_{1}=3+i$, $z_{3}=0$ and $z_{4}=3$ are given as:
\begin{eqnarray}
&&\hspace{5cm} Q_{1}(z,t)=-2e^{6iz}\frac{G_{1}}{F}, \quad Q_{2}(z,t)=-e^{6iz}\frac{G_{2}}{F},\nonumber\\
&&G_{1}=36 z_2^4 t^4+288 \sqrt{3} z_2^3 t^4-144 \sqrt{3} z_2^2 t^4+2952 z_2^2 t^4+144 \sqrt{3} t^4+4896 \sqrt{3} z_2 t^4-1728 z_2 t^4\nonumber\\
&&+16020 t^4+24 \sqrt{3} z_2^4 t^3+288 z_2^3 t^3+384 \sqrt{3} z_2^2 t^3-144 z_2^2 t^3+456 \sqrt{3} t^3-288 z_2 t^3+2736 t^3\nonumber\\
&&+864 z^2 z_2^4 t^2-144 i z z_2^4 t^2+12 z_2^4 t^2-1152 i \sqrt{3} z z_2^3 t^2+6912 z^2 \sqrt{3} z_2^3 t^2-36 \sqrt{3} z_2^3 t^2+384480 z^2 t^2\nonumber\\
&&-3456 \sqrt{3} z^2 z_2^2 t^2+70848 z^2 z_2^2 t^2-11808 i z z_2^2 t^2-(864-576 i) \sqrt{3} z z_2^2 t^2+72 i \sqrt{3} z_2^2 t^2-456 z_2^2 t^2\nonumber\\
&&+(444-1296 i) t^2+(15552-64080 i) z t^2-(864+576 i) \sqrt{3} z t^2+3456 z^2 \sqrt{3} t^2+72 i \sqrt{3} t^2\nonumber\\
&&-41472 z^2 z_2 t^2+6912 i z z_2 t^2+117504 z^2 \sqrt{3} z_2 t^2-(612+288 i) \sqrt{3} z_2 t^2+(3456-19584 i) z \sqrt{3} z_2 t^2\nonumber\\
&&+216 z_2 t^2-48 i \sqrt{3} z z_2^4 t+288 z^2 \sqrt{3} z_2^4 t+3456 z^2 z_2^3 t-576 i z z_2^3 t-36 z_2^3 t+32832 z^2 t-1728 z^2 z_2^2 t\nonumber\\
&&+288 i z z_2^2 t-768 i \sqrt{3} z z_2^2 t+4608 z^2 \sqrt{3} z_2^2 t-5472 i z t-912 i \sqrt{3} z t+5472 z^2 \sqrt{3} t-3456 z^2 z_2 t\nonumber\\
&&+(36-216 i) z_2 t+(2592+576 i) z z_2 t-216 i \sqrt{3} z_2 t+2592 z \sqrt{3} z_2 t+2306880 z^4+5184 z^4 z_2^4\nonumber\\
&&-1728 i z^3 z_2^4-24 i z z_2^4-z_2^4+(186624-768960 i) z^3-(10368+6912 i) \sqrt{3} z^3-13824 i \sqrt{3} z^3 z_2^3\nonumber\\
&&-720 \sqrt{3} z^2 z_2^3-72 i \sqrt{3} z z_2^3+41472 z^4 \sqrt{3} z_2^3-6 \sqrt{3} z_2^3-(58752+46656 i) z^2-(576-2592 i) \sqrt{3} z^2\nonumber\\
&&-20736 \sqrt{3} z^4 z_2^2+425088 z^4 z_2^2-141696 i z^3 z_2^2-(10368-6912 i) \sqrt{3} z^3 z_2^2-6912 z^2 z_2^2\nonumber\\
&&-816 i z z_2^2+(576+2592 i) z^2 \sqrt{3} z_2^2+144 z \sqrt{3} z_2^2-34 z_2^2-(2592+888 i) z+20736 z^4 \sqrt{3}+144 z \sqrt{3}\nonumber\\
&&-248832 z^4 z_2+82944 i z^3 z_2-(12240+10368 i) \sqrt{3} z^2 z_2+4320 z^2 z_2+432 i z z_2-(144+1224 i) \sqrt{3} z z_2\nonumber\\
&&+705024 z^4 \sqrt{3} z_2+(41472-235008 i) z^3 \sqrt{3} z_2-(102+36 i) \sqrt{3} z_2-1,\nonumber
\end{eqnarray}
\begin{eqnarray}
&&F=36 t^4 z_2^4+288 \sqrt{3} t^4 z_2^3-144 \sqrt{3} t^4 z_2^2+2952 t^4 z_2^2+4896 \sqrt{3} t^4 z_2-1728 t^4 z_2+144 \sqrt{3} t^4+16020 t^4\nonumber\\
&&+24 \sqrt{3} t^3 z_2^4+288 t^3 z_2^3+384 \sqrt{3} t^3 z_2^2-144 t^3 z_2^2-288 t^3 z_2+456 \sqrt{3} t^3+2736 t^3+864 t^2 z^2 z_2^4\nonumber\\
&&+6912 \sqrt{3} t^2 z^2 z_2^3+384480 t^2 z^2-3456 \sqrt{3} t^2 z^2 z_2^2+70848 t^2 z^2 z_2^2+3456 \sqrt{3} t^2 z^2-41472 t^2 z^2 z_2\nonumber\\
&&+117504 \sqrt{3} t^2 z^2 z_2+24 t^2 z_2^4+48 \sqrt{3} t^2 z_2^3-864 \sqrt{3} t^2 z z_2^2-24 \sqrt{3} t^2 z_2^2+384 t^2 z_2^2\nonumber\\
&&-864 \sqrt{3} t^2 z+15552 t^2 z+3456 \sqrt{3} t^2 z z_2+816 \sqrt{3} t^2 z_2+24 \sqrt{3} t^2+4344 t^2+288 \sqrt{3} t z^2 z_2^4+3456 t z^2 z_2^3\nonumber\\
&&+32832 t z^2-1728 t z^2 z_2^2+4608 \sqrt{3} t z^2 z_2^2+5472 \sqrt{3} t z^2-3456 t z^2 z_2+4 \sqrt{3} t z_2^4-864 t z z_2^2+64 \sqrt{3} t z_2^2\nonumber\\
&&-16416 t z+3456 t z z_2+76 \sqrt{3} t+2306880 z^4+5184 z^4 z_2^4+41472 \sqrt{3} z^4 z_2^3-20736 \sqrt{3} z^4 z_2^2+425088 z^4 z_2^2\nonumber\\
&&+20736 \sqrt{3} z^4-248832 z^4 z_2+705024 \sqrt{3} z^4 z_2-10368 \sqrt{3} z^3+186624 z^3-10368 \sqrt{3} z^3 z_2^2+41472 \sqrt{3} z^3 z_2\nonumber\\
&&+144 z^2 z_2^4+576 \sqrt{3} z^2 z_2^3+22608 z^2-288 \sqrt{3} z^2 z_2^2+6624 z^2 z_2^2+288 \sqrt{3} z^2\nonumber\\
&&-6912 z^2 z_2+9792 \sqrt{3} z^2 z_2+z_2^4-144 \sqrt{3} z z_2^2+34 z_2^2-144 \sqrt{3} z+576 \sqrt{3} z z_2+1.\nonumber
\end{eqnarray}
\begin{eqnarray}
&&G_{2}=36 i z_2^4 t^4+288 i \sqrt{3} z_2^3 t^4+2952 i z_2^2 t^4-144 i \sqrt{3} z_2^2 t^4+16020 i t^4+144 i \sqrt{3} t^4-1728 i z_2 t^4\nonumber\\
&&+4896 i \sqrt{3} z_2 t^4+24 i \sqrt{3} z_2^4 t^3+288 i z_2^3 t^3-144 i z_2^2 t^3+384 i \sqrt{3} z_2^2 t^3+2736 i t^3+456 i \sqrt{3} t^3-288 i z_2 t^3\nonumber\\
&&+864 i z^2 z_2^4 t^2+12 i z_2^4 t^2+144 z z_2^4 t^2+6912 i z^2 \sqrt{3} z_2^3 t^2+1152 z \sqrt{3} z_2^3 t^2+384480 i z^2 t^2+70848 i z^2 z_2^2 t^2\nonumber\\
&&-3456 i \sqrt{3} z^2 z_2^2 t^2-24 i z_2^2 t^2-(576+864 i) \sqrt{3} z z_2^2 t^2+11808 z z_2^2 t^2-(72+72 i) \sqrt{3} z_2^2 t^2+(1296+4764 i) t^2\nonumber\\
&&+(64080+15552 i) z t^2+3456 i z^2 \sqrt{3} t^2-(72-72 i) \sqrt{3} t^2+(576-864 i) z \sqrt{3} t^2-41472 i z^2 z_2 t^2-864 i z_2 t^2\nonumber\\
&&-6912 z z_2 t^2+117504 i z^2 \sqrt{3} z_2 t^2+(19584+3456 i) z \sqrt{3} z_2 t^2+288 \sqrt{3} z_2 t^2+288 i z^2 \sqrt{3} z_2^4 t\nonumber\\
&&+48 z \sqrt{3} z_2^4 t+3456 i z^2 z_2^3 t+576 z z_2^3 t+32832 i z^2 t-1728 i z^2 z_2^2 t+(216-72 i) z_2^2 t-(288-2592 i) z z_2^2 t\nonumber\\
&&+4608 i z^2 \sqrt{3} z_2^2 t+768 z \sqrt{3} z_2^2 t+(4104+1368 i) t+(5472+49248 i) z t+5472 i z^2 \sqrt{3} t+912 z \sqrt{3} t\nonumber\\
&&-3456 i z^2 z_2 t-576 z z_2 t+10368 i z \sqrt{3} z_2 t+864 \sqrt{3} z_2 t+2306880 i z^4+5184 i z^4 z_2^4+1728 z^3 z_2^4-i z_2^4\nonumber\\
&&+24 z z_2^4+(768960+186624 i) z^3-1152 i \sqrt{3} z^2 z_2^3+41472 i z^4 \sqrt{3} z_2^3+13824 z^3 \sqrt{3} z_2^3+(46656-110592 i) z^2\nonumber\\
&&-(2592+1440 i) \sqrt{3} z^2+425088 i z^4 z_2^2-20736 i \sqrt{3} z^4 z_2^2-(6912+10368 i) \sqrt{3} z^3 z_2^2\nonumber\\
&&+141696 z^3 z_2^2-12096 i z^2 z_2^2-(2592-1440 i) \sqrt{3} z^2 z_2^2-34 i z_2^2-48 z z_2^2+(36-12 i) \sqrt{3} z_2^2\nonumber\\
&&+(144+576 i) z \sqrt{3} z_2^2-i-(7752+2592 i) z-(144-576 i) \sqrt{3} z+20736 i z^4 \sqrt{3}+(6912-10368 i) z^3 \sqrt{3}\nonumber\\
&&+(36+12 i) \sqrt{3}-248832 i z^4 z_2-82944 z^3 z_2+17280 i z^2 z_2-576 i \sqrt{3} z z_2+1728 z z_2+705024 i z^4 \sqrt{3} z_2\nonumber\\
&&+(235008+41472 i) z^3 \sqrt{3} z_2+(10368-19584 i) z^2 \sqrt{3} z_2,\nonumber
\end{eqnarray}

\end{document}